\documentclass{article}

\usepackage[numbers]{natbib}  
\usepackage{graphicx}          
\usepackage{amsmath}           
\usepackage{siunitx}           
\usepackage{float}             
\usepackage{caption}           
\usepackage{subcaption}        
\usepackage{amssymb}           
\usepackage{mathrsfs}          
\usepackage{physics}           
\usepackage{xcolor}            
\usepackage{fancyhdr}          
\usepackage{afterpage}         
\usepackage{comment}           

\newenvironment{changemargin}[2]{%
  \begin{list}{}{\leftmargin#1\rightmargin#2}%
  \item[]}{\end{list}}

\usepackage{hyperref}

\usepackage[countmax]{subfloat}

\usepackage[labelfont=bf]{caption}  
\usepackage{subfloat}               
\usepackage{mathtools}              
\usepackage{mwe}                    
\usepackage{amsthm}                 
\usepackage{imakeidx}               
\usepackage{authblk}                
\usepackage[font={small}]{caption}  

\usepackage[a4paper, total={6in, 8in}]{geometry}

\theoremstyle{plain}

\theoremstyle{definition}
\newtheorem{definition}{Definition}

\theoremstyle{remark}

\title{Weak Particle Presence}
\author{Bethany Terris\\ Université Paris-Saclay, CEA, LARSIM, 91191 Gif-sur-Yvette, France}
\date{bethany.terris@cea.fr, https://orcid.org/0000-0001-8786-7069}

\begin{document}

\maketitle

\begin{abstract}
    The concept of presence has been extensively explored in philosophy, yet the notion of particle presence within quantum theory remains under-examined. In this article, we explore particle presence through an analysis of a paradox arising from weak measurements. We show that the classical intuition about particle presence involves an erroneous logical combination of propositions from single-time weak values, leading to inconsistencies that result in the deduction of discontinuous trajectories. Instead, we argue that by treating presence as a property defined across time by measuring sequential weak values, the discontinuity paradox is resolved, providing a coherent, non-classical account of particle presence. We discuss some advantages and drawbacks of this account, and consider applications to other cases of trajectory discontinuity.
\end{abstract}

\textbf{Keywords}
Particle presence, Weak measurement, Sequential weak values, Quantum paradoxes

\newpage


\section{Introduction} \label{Sec:Intro}

The notion of presence has long been subject to philosophical and metaphysical inquiry, encompassing foundational analyses by the likes of Aquinas, Kant, and Hume, who each develop methodical accounts of presence, grounded in their broader metaphysical frameworks \cite{aquinas1968being,hume2000treatise,Kant1929-KANCOP-17}. Critiques like Derrida’s deconstruction reconceptualise presence in relation to absence \cite{derrida1973speech,Derrida1976-DEROG,Derrida1978-DERWAD-3,derrida1982margins}, while Heidegger’s exploration of presence centres on the concept of `Being', examining how our understanding of presence is intrinsically linked to time \cite{heidegger1962heidegger}. While there has been some analysis of presence within quantum theory \cite{arvidsson2018quantum}, particularly concerning the identity and individuality of particles \cite{Cote2013-GILTMA-3,dieks2023emergence,falkenburg2007particle,Goyal2019-GOYPAN,krause2023identity,Redhead1991-REDPPL,Schrodinger1950-SCHWIA-29}, there has been comparatively less philosophical attention awarded to presence within quantum theory. In classical physics, particle presence is understood through a single-time approach, which attempts to pinpoint the position of a particle at a discrete moment in time \cite{hance2023weak}. Single-time measurements are then logically combined to deduce the particle's trajectory through spacetime. However, applying this single-time intuition to quantum mechanics leads to paradoxes involving discontinuous particle trajectories, particularly when weak values are involved.

A single-time weak value can be found by conditioning on the pre- and post-selected states of a quantum system. The weak value $A_W$ is defined for any operator $\hat{A}$, pre-selected in the initial state $\ket{\psi}$ and post-selected in the final state $\ket{\phi}$ \cite{aharonov2008two}:

\begin{equation}
    A_W=\frac{\bra{\phi}\hat{A}\ket{\psi}}{\braket{\phi}{\psi}}
    \label{eq:weakvaluetsvf}
\end{equation}

Weak values differ from conventional quantum measurement procedures as they can exceed the eigenvalue range of $\hat{A}$ \cite{aharonov1988result}. They provide a way to extract information about quantum systems with minimal disturbance, offering insights that traditional projective measurements cannot, which irreversibly alter the system, placing a limit on the amount of information that can be extracted. This approach opens doors to observing phenomena that are otherwise inaccessible. For example, weak values can provide insight into the path of a quantum particle, with negligible disturbance. When single-time weak values are used to indicate the presence of a single particle in a discrete spatiotemporal location, discontinuous particle trajectories are deduced, challenging the classical assumption of a continuous spacetime manifold \cite{aharonov2013quantum,danan2013asking,vaidman2013past,vaidman2014tracing,aharonov2017case,das2020can,liu2020experimental,aharonov2021dynamical,okamoto2023experimentally}.

This article sheds light on this problem by arguing that single-time weak values are insufficient for deducing particle trajectories. This is due to the logical inconsistency that arises from the combination of single-time propositions about the particle's position. We argue that this method treats presence as primarily being defined at discrete times, thereby not addressing the persistence of the particle in spacetime.

While some may be willing to accept discontinuity as an inevitable consequence of quantum theory, this article seeks to explore whether it is possible to preserve classical intuitions within the framework of weak values, and what the cost of doing so is. Hence, our argument attempts to reconcile the following two assumptions:

\hfill

\textbf{Assumption 1:} Particle trajectories are continuous in spacetime.

\hfill

\textbf{Assumption 2:} A weak value represents a genuine property of an individual system.

\hfill

The discontinuity paradoxes create a tension between our two assumptions, whereby we cannot maintain that trajectories are continuous whilst also taking weak values seriously. We reconcile these assumptions by restricting claims of particle presence to sequential weak values (SWVs) which explore the weak value of operators $A_1...A_N$ across multiple sequential times. The realisation of SWVs has been developed in \cite{mitchison2007sequential}, and they are measured using the following equation which again conditions on the pre- and post-selections of the system, but now multiple sequential times are considered:

\begin{equation} \label{SWV eqn}
\langle A_n,...,A_1 \rangle _W=\frac{\bra{\phi}\hat{A_n}...\hat{A_1}\ket{\psi}}{\braket{\phi}{\psi}}
\end{equation}

We assert the following criterion which must be met in order to denote particle presence:

\hfill

\begin{changemargin}{0.5cm}{0.5cm} 

A pair of overlapping $n$-tuple sequential weak values of the spatial projection operator, each yielding non-zero results, permit a claim of particle presence across the locations and duration described by the sequential weak values.

\end{changemargin}

\hfill

We argue that SWVs which meet this criterion form a consistent story about the history of a particle's trajectory. The measurement of an SWV asks a compound question regarding the trajectory across a time duration, and avoids the erroneous logical combination of inconsistent single-time propositions. As a result, SWVs address the persistence of the particle in spacetime, recovering trajectory continuity and reconciling Assumptions 1 and 2.

The article is organised as follows. Section \ref{Sec:cont} justifies Assumption 1 and develops the classical, single-time account of particle presence. Section \ref{Sec:disc} begins by justifying Assumption 2, then discusses a discontinuity paradox that arises from the measurement of single-time weak values. We argue that this paradox can be resolved by measuring SWVs which form a consistent set of histories, and assert the presence criterion which determines the conditions under which SWVs ought to assign particle presence. In Sec \ref{Sec:acrosspres}, the across-time account of particle presence is developed and comparisons are drawn with classical particle presence. Additionally, some advantages and drawbacks of our account are discussed, including an application of this account to resolve another discontinuity paradox. Sec \ref{further} discusses some possible further research directions related to our argument, and finally, Sec \ref{Sec:conc} draws conclusions, noting the significance of our result in improving upon a metaphysical understanding of presence.

\section{Continuity and Classical Particle Presence} \label{Sec:cont}

Classical physics assumes a continuous spacetime metric, which corresponds to a three-dimensional topological manifold with homeomorphic transformations of the form $f: X \rightarrow Y$, where the function is a continuous, one-to-one map between two regions, $X$ and $Y$ \cite{freedman1990topology}. The inverse function $f^{-1}$ is also continuous, meaning that after undergoing homeomorphisms, certain properties remain invariant, such as dimensionality and connectedness \cite{lefschetz1923continuous,poincare1895analysis}. This results in a continuous spacetime, which is considered to be a background upon which events unfold and entities traverse. As particle trajectories are assumed to be trajectories through a continuous background spacetime, the trajectories themselves are assumed to be continuous in spacetime in classical physics \cite{maudlin2012philosophy}.

In quantum mechanics, the assumption of continuous particle trajectories is challenged by numerous phenomena and even the framework of quantum theory, beginning with Bohr’s atomic theory which rejected the idea of well-defined particle trajectories independent of measurement \cite{bohr1913constitution}. Heisenberg’s matrix mechanics abandoned the notion of particle trajectories altogether, focusing on observable quantities rather than representations of particle behaviour \cite{heisenberg1925quantum}. The uncertainty relations preclude continuous trajectories by treating quantum measurements as inherently discrete, meaning that they cannot approximate continuity, unlike in classical physics. Additionally, in quantum theory, elementary particles are largely agreed to be point-like, such that they are dimensionless and lack substructure and spatial extension \cite{hobson2013there}, making it difficult to describe the movement of point-like particles through spacetime.

The definition of `particle' is obscured, which further complicates how we may speak about particle trajectories. First, particles are considered indistinguishable, meaning it is impossible to uniquely identify or differentiate one fermion from another based solely on their intrinsic properties. This inability to correctly label particles makes it difficult to track their trajectories, challenging Assumption 1. However, this indistinguishability does not pose a threat when attempting to track a particle's spatiotemporal evolution, as this evolution helps us to distinguish between particles. Continuous trajectories thus provide a valuable tool for describing the dynamical evolution of systems over time, and help in distinguishing what may otherwise be indistinguishable particles.

Second, the collective behaviour of a group of particles can be approximated by treated the group as a single `quasiparticle', further complicating the definition of `particle' \cite{2003aess.book.....K,landau1957theory}. Electron holes are an example of quasiparticles which are posited to denote the positive charge left by the lack of an electron where one could be in an atomic lattice, particularly in the valence band of a superconductor \cite{ashcroft1976solid,shockley1950electrons}. With an obscured definition of particles, it is unclear how we should speak of particle trajectories. However, the existence of quasiparticles does not eliminate the utility of continuous trajectories. Instead, it reframes the discussion: quasiparticles are effective theoretical constructs that help model collective behaviours and interactions within a system. The trajectories of these quasiparticles could still be treated as continuous within their framework, offering insights into the system's evolution and dynamics.

Another critique of Assumption 1 comes from phenomena arising from quantum field theory (QFT). QFT provides a perspective of particles that diverges from the classical view, defining particles as excitations of fields \cite{becker2006string,neumann1955mathematical}. Additionally, there is no strict boundary separating particles from non-particles. In QFT, the concept of particle trajectories is typically less well-defined than in standard non-relativistic quantum mechanics. Particles can undergo quantum jumps, annihilation, pair production, and other discontinuous events that challenge Assumption 1. Discontinuous trajectories may be necessary to account for phenomena like pair production and annihilation in QFT, but the focus of this article is on trajectories within the non-relativistic quantum mechanical framework. This focus allows us to set aside cases of possible discontinuity arising from QFT, without impacting the scope of the discussion.

In response to these critiques, one may ask: why do we make Assumption 1 if this assumption is frequently challenged in quantum theory? Despite these critiques, continuous trajectories are intuitive, and extending them to quantum theory provides a way to speak about particles that is in alignment with our classical intuition. Our intention in this article is to analyse whether it is possible to extend this intuition to quantum theory, and what the cost of doing so is. This is not to say that the continuous perspective is the only correct one, but that adopting this perspective provides a straightforward way to understand particles and their movement in spacetime, which could help us to develop an understanding of particles in quantum theory. Additionally, despite the challenge posed by the quantum framework to continuous trajectories, several interpretations of quantum mechanics agree with Assumption 1 (e.g. consistent histories, many worlds). Cases of discontinuity pose a threat to these interpretations, but if a framework for recovering continuity is proposed, it is possible to save them.


The continuity referred to here is defined in terms of both its spatial and temporal components \cite{skow2007makes}. Spatial continuity implies that particles move on a continuous path across space, and cannot access spatiotemporal locations that are space-like separated from the particle's initial location due to the restriction from special relativity \cite{einstein1905electrodynamics}. For instance, for a particle on a trajectory between two locations $A$ and $B$ which are spatially distant from each other, the particle must move across a series of locations that are spatially adjacent and that thus form a continuous chain of spatially adjacent locations from $A$ to $B$. The particle cannot instantaneously jump straight from $A$ to $B$ without moving across this continuous chain of locations.

Meanwhile, temporal continuity implies that the particle follows a continuous path in time. This means that for a trajectory across the time duration $[t_i, t_f]$, the particle must be able to be located in spacetime throughout this entire duration. During the particle's trajectory, the particle must not `jump' through time, i.e. the particle should not disappear from spacetime after $t_i$ and re-appear at $t_f$.

Continuous particle trajectories are typically considered to be a condition for particle presence, particularly within classical domains. So, continuity provides an understanding of what it means to be a particle, and for that particle to be `present’. The classical conception of particle presence is characterised by the following HRL conditions \cite{hance2023weak}:

\begin{itemize}
\setlength\itemsep{0.00005em}
    \item[(i)] Each particle has a position in space at all times.
    \item[(ii)] A single particle cannot be on more than one spacetime trajectory simultaneously.
    \item[(iii)] A particle's trajectory is continuous such that the particle cannot follow a trajectory from one position to another without passing through the positions in between.
    \item[(iv)] A particle only interacts with the objects or fields local to its spacetime position.
    \item[(v)] If a particle is on a trajectory which is located within some region at a given time, then the particle is also located within that region at that time.
    \item[(vi)] If the property belonging to a particle is in a particular spacetime position, then the particle must also be in that spacetime position.
\end{itemize}

The authors, HRL, argue that in order to make a claim of particle presence, each of the conditions (i)---(vi) must be met. However, there are several cases where at least one of these conditions are violated in quantum mechanics. For example, particles can be in a superposition of multiple position states \cite{dirac1981principles}. This could lead to differing interpretations of the particle’s physical state: one might argue that the particle lacks a definite spatial location, potentially violating condition (i), or that the particle is simultaneously located in multiple positions, violating condition (ii). However, these are not the only possible interpretations of position states. Following Dirac’s formalism \cite{dirac1981principles}, a position state refers to the state vector in Hilbert space, which is a mathematical representation of the system that does not necessarily describe the physical state of the particle. The relationship between the state vector and the physical state depends on the interpretation of quantum mechanics, and if one assumes no link, then the violation of these conditions does not arise. In this article, we adopt the assumption that the state vector represents the particle’s physical position, due to the focus on presence.

Condition (iii) may be violated by the phenomenon of quantum tunnelling \cite{gurney1929quantum,hund1927linienspektren}. The possibility for a particle to tunnel through a potential barrier indicates a discontinuous particle trajectory in space. The particle tunnels from one side of the barrier to the other side, without passing through the spatial positions in between and thus is at odds with condition (iii). This violation is similarly interpretation-dependent, for example in the path integral formulation of quantum mechanics, the particle takes all possible paths simultaneously, including through the tunnelling region. However, this interpretation does not imply discontinuity; instead, it integrates over all continuous paths, including those with low classical probability, to calculate the overall amplitude. The concept of discontinuity arises only in terms of classical intuition, not in the mathematical description of the particle’s behaviour.

Quantum entanglement describes a non-local correlation between particles such that when a pair of particles are separated by large distances, the state of one particle is dependent upon the state of its pair \cite{einstein1935can}. As condition (iv) requests locality, this condition is in contention with quantum entanglement which suggests non-local correlations when interpreted in this way. However, in many quantum mechanical interpretations (many worlds, consistent histories, relational quantum mechanics),  quantum entanglement does not necessarily imply a direct violation of locality but rather suggests that quantum systems do not adhere to classical intuitions about separability and independence. In these interpretations, the correlations observed in entangled systems are consistent with the locality condition (iv) in the sense that they do not imply instantaneous signalling or influence between distant particles. This particular interpretation of entanglement makes it possible to avoid the violation of condition (iv), while ontic interpretations of entanglement lead to the violation.

An example of the violation of condition (v) is the weak trace approach to the nested interferometer paradox \cite{vaidman2013past,vaidman2014tracing}. This approach uses two-state vector formalism to argue that the path followed by the particle is defined by the overlap between the states evolving both forward in time from the pre-selection and backward in time from the post-selection. But there are regions occupied by the forward or backward evolving states alone that are argued to not be occupied by the particle. If one wishes to assign realism to the forward and backward evolving states then this leads to a contradiction with HRL condition (v) as the particle is not located within the region occupied by its forward or backward evolving trajectories. Again, avoiding an ontic interpretation of the quantum state means that this violation can be avoided.

Finally, HRL condition (vi) is violated by the quantum Cheshire cat effect \cite{aharonov2013quantum,das2020can,ibnouhsein2012twin,kim2021observing,liu2020experimental}. This phenomenon utilises weak measurements to investigate the trajectory of a particle through an interferometer with two distinct paths $A$ and $B$. The results of the experiment suggest that the particle is restricted to a trajectory along path $A$, meanwhile the spin property of the particle follows a trajectory along path $B$. So, the particle and its property occupy different spatial positions at the same time, violating condition (vi). Perhaps it is possible to interpret the effect as a reflection of the relational dynamics between the particle, its property, and the measurement apparatus, rather than an actual separation in physical space. The violation of condition (vi) here is once again interpretation-dependent, but the threat to the condition should be taken seriously.

The classical conditions for particle presence are formulated in a way that assumes a single-time intuition, whereby presence of an individual particle is claimed in a position at a single time. For example, a proposition about particle presence that assumes the single-time intuition is of the form, `The particle is in location $x$ at time $t$'.

The single-time intuition about particle presence is extended to the discussion of particle trajectories. Classically, for $N$ times $t_i...t_N$, the particle has a position at each of the $N$ times. A proposition regarding particle presence can then be made in each of these positions, at each of these discrete times. The particle's trajectory is then deduced from a logical combination of several single-time propositions of particle presence each made at successive times.

By a `logical combination' we mean the combination of a number of propositions in order to form a more complex proposition, using the logical AND operator, $\land$. For example, combining the propositions `The particle is in location $X$ at time $t_1$' and `The particle is in location $Y$ at time $t_2$' involves applying the $\land$ operator, such that the complex proposition `The particle is in location $X$ at time $t_1$ and the particle is in location $Y$ at time $t_2$' returns the truth value TRUE iff each of the single-time propositions have truth value TRUE. It is expected that if the complex proposition returns TRUE, the single-time propositions are consistent such that there is no contradiction between the propositions. If the truth value of the complex proposition is FALSE, then this suggests inconsistency and thus discontinuity. But if the complex truth value is TRUE, the consistency of the single-time propositions suggests that there is no discontinuity, meaning that the particle follows the continuous trajectory $[X,Y]$ over $[t_1,t_2]$.

The mention of `at all times' in HRL condition (i) signifies an account where time is divisible into single, discrete moments. The use of this phrase in condition (i) alludes that multiple single-time propositions can be logically combined to indicate continuous particle trajectories. Thus condition (i) assumes a single-time intuition about particle presence. Both conditions (i) and (iii) together refer to the continuity of the particle's trajectory in spacetime. Condition (i) considers the continuity of the particle's trajectory in time, as it requires that the particle remains located in space at every discrete time. A violation of this condition would suggest the disappearance of the particle at some point in time, and so the particle's trajectory would be discontinuous in time. Meanwhile, condition (iii) considers the continuity of the particle's trajectory in space as it requires that the particle passes through a continuous chain of spatial positions. Condition (iii) makes this argument by assuming that single-time claims of presence can be combined to indicate particle trajectories which are spatially continuous, and thus this condition also assumes the single-time intuition.

The remainder of the conditions also hint at single-time intuition. Condition (ii) refers to simultaneity, which is a single-time notion as simultaneity generally refers to two or more events occurring at the same discrete point in time. Conditions (iv) and (vi) involve a notion of `being in a position' which refers to the particle's position at a single point in time. Thus, these conditions have also been formulated with the assumption of single-time intuition. Condition (v) considers the trajectory of the particle, but at a single moment in time, for which the particle is required to be located within the same region as that which the trajectory occupies. As a result, this condition also assumes the single-time intuition.

So, each of the HRL conditions for classical particle presence assume the single-time intuition. In the classical account of presence, single-time propositions about presence are primary, and propositions regarding trajectories are secondary as they are deduced from a logical combination of multiple single-time propositions. In this sense, classical physics assumes that particle trajectories are derivative of single-time propositions regarding the particle's position.

\begin{figure}[H]
    \centering
    \includegraphics[scale=0.2]{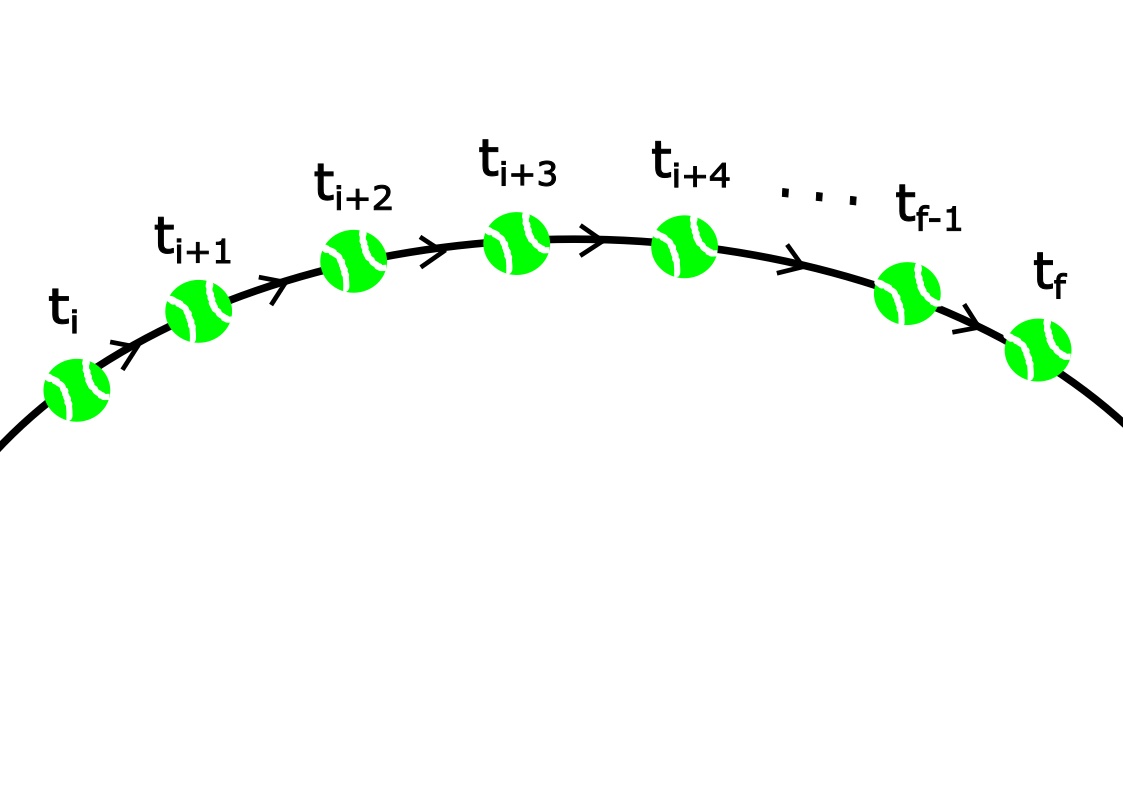}
    \caption{The trajectory (black arrow) of a tennis ball across a time duration $[t_i,t_f]$, as told by the single-time positions (green balls)}
    \label{fig:tennis}
\end{figure}

The single-time intuition is present throughout classical physics, not just in the specific case of particle presence. For example, the same intuition is assumed when considering the dynamics of macroscopic objects. In classical mechanics, the trajectory of a tennis ball being hit across a court can be logically deduced from multiple propositions made about the position of the tennis ball at discrete times, similarly to particle trajectories. The propositions we refer to are usually derived from measurements, and so it is the measurement procedure that assumes this single-time intuition. Consider Fig \ref{fig:tennis} which demonstrates the trajectory of a tennis ball experiencing projectile motion. The positions of the tennis ball at successive individual times $t_i...t_f$ that are deduced from position measurements are logically combined to form a story about a continuous trajectory as demonstrated by the black arrow.

Single-time propositions over a greater number of times provide more information which can be used to build a more complete picture of the trajectory. For example, a logical combination of propositions regarding the position of the tennis ball at each discrete time displayed in Fig \ref{fig:tennis} provides a more accurate deduction of the trajectory than the logical combination of propositions just at times $t_i$, $t_{i+3}$, and $t_f$, even though both cases concern the same duration $[t_i,t_f]$. Classically, particle presence is treated in the same way. A larger number of successive single-time propositions regarding the particle's spacetime position provides a more complete story about the trajectory than fewer single-time propositions across the same time duration, as they provide more information. Ultimately, trajectories deduced from a logical combination of single-time propositions are conjectures, based on incomplete information regarding the particle's position over the time duration. This method of deducing trajectories leads to problems, including an apparent discontinuity in spacetime \cite{aharonov2017case}.

\section{Discontinuity} \label{Sec:disc}
\subsection{A discontinuity paradox} \label{Subsec:paradox}

A number of paradoxes involving discontinuous particle trajectories arise from allowing single-time weak values to identify the positions of a particle at multiple discrete times \cite{aharonov2013quantum,aharonov2017case,aharonov2021dynamical,das2020can,liu2020experimental,vaidman2013past,vaidman2014tracing}. One way of avoiding these paradoxes is to reject Assumption 1, arguing that continuous trajectories are not necessary in quantum theory. We have explored this reasoning in Sec \ref{Sec:cont}, arguing that our intention is to see whether it is possible to extend this assumption to quantum theory. While we acknowledge that this is not the only possible interpretation, we also note the pragmatism in maintaining continuity, given its potential to identify and describe the behaviour of quantum systems across time.

Another route out of these paradoxes is to reject Assumption 2, arguing that weak values do not represent a property of an individual system. Arguments of this type have previously been explored, claiming that weak values are purely statistical averages that arise from weak measurements performed on a large ensemble \cite{alves2017achieving,ferrie2014result,ferrie2014weak,ferrie2015ferrie,gross2015novelty,knee2014amplification,knee2016weak,reznik2023photons}. However, a counterargument operationally relates weak values to eigenvalues, reasoning that they can be understood beyond their statistical features and thus ought to be recognised as representing a property of an individual system \cite{vaidman2017weak,vaidman2017beyond,matzkin2019weak}. It is shown that the weak value of a system governs the system's interactions similar to how an eigenvalue does. More specifically, given a system that is in an eigenstate of an observable $A$, a measurement on $A$ will yield the eigenvalue related to that eigenstate, so the eigenvalue is dependent on the eigenstate. Meanwhile, the contextuality of the weak value arises as a result of its dependence on the pre- and post-selected states. This dependency governs the outcome of a (weak) measurement interaction on the system's observable, similar to how the dependency on the eigenstate governs the eigenvalue in a standard measurement. Whilst there are similarities between weak values and eigenvalues, it is important to note that weak values exhibit stronger contextuality than eigenvalues.

An expectation value is commonly interpreted as the statistical average of all possible measurement outcomes on an ensemble of systems. Meanwhile, in other interpretations like QBism, the expectation value quantifies an agent's beliefs about what they will observe based on their prior knowledge and the probabilistic rules of quantum mechanics. In both of these interpretations, the expectation value does not tell us the outcome of a single measurement. Rather, the eigenvalue plays this role, while the expectation value represents a probabilistic result, or a subjective belief. It is for this reason that weak values can be interpreted beyond their statistical features, as they govern the outcome of a (weak) measurement interaction on a system, conditioned by the pre- and post-selected states, much like how an eigenvalue is determined by the eigenstate of an observable in a standard quantum measurement. 

Additionally, empirical confirmation plays a role in interpreting weak values as having eigenvalues as their classical analogue. A range of experimental evidence suggests that weak values are appropriately understood as representing a property of an individual system, and thus ought to be understood as such rather than as statistical averages \cite{korotkov2001continuous,lundeen2012procedure,strubi2013measuring,xu2013phase,zhou2013weak,jayaswal2014observing,magana2014amplification,lyons2015power,zhang2015precision,hallaji2017weak}.

We do not intend to add to the debate regarding the interpretation of weak values, but we believe we have demonstrated that Assumption 2 is justified. Our objective is to consider whether we can maintain continuity if we are to take weak values seriously. We thus make Assumption 2, and we argue that the discontinuity paradoxes presented by the measurement of single-time weak values leads to tension between Assumptions 1 and 2, such that we cannot simultaneously believe that the single-time weak values represent the positions of an individual particle whilst also maintaining continuous particle trajectories. In the following, we discuss one of these paradoxes and propose a solution to the discontinuity problem, thus bringing about harmony between our assumptions, eliciting a departure from the classical intuition about particle presence as a cost of this solution.

\begin{figure}
    \centering
    \includegraphics[scale=0.25]{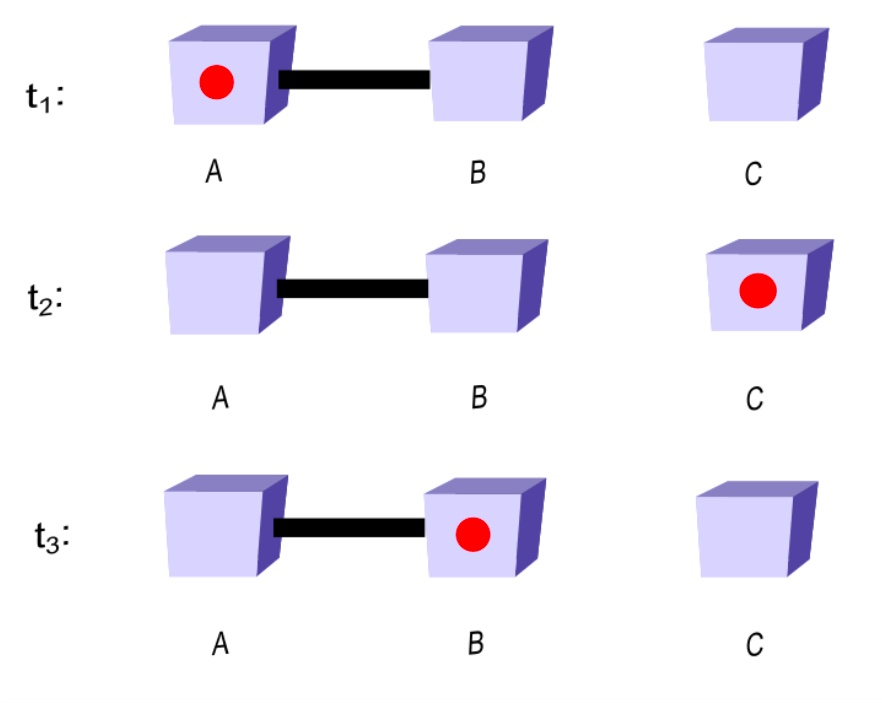}
    \caption{Discontinuous trajectory from single-time weak values: The particle follows a discontinuous trajectory in either space or time from boxes $A \rightarrow C \rightarrow B$ over $[t_1,t_3]$. The discontinuity in space arises as the particle jumps from box $A$ to $C$ without passing through $B$. Ignoring $\langle C(t_2) \rangle _W$ helps to avoid spatial discontinuity, but implies temporal discontinuity as this signifies that the particle disappears from $A$ after $t_1$ and re-appears in $B$ at $t_3$. So here, the particle is not located in spacetime at $t_2$.}
    \label{fig:dnr}
\end{figure}

The gendankenexperiment is comprised of a particle that is pre- and post-selected as superposed over three boxes, $A$, $B$ and $C$ as seen in Fig \ref{fig:dnr} \cite{aharonov2017case}. Tunnelling is permitted between boxes $A$ and $B$ only. In order to investigate the position of the particle throughout the duration of the experiment, single-time weak values of the spatial projection operator are measured in each of the three boxes at three successive times, $t_1$, $t_2$ and $t_3$. It is the measurement of these weak values that leads to the discontinuous trajectories.

The following single-time weak values are measured in \cite{aharonov2017case}:

\begin{equation} \label{eq:At1}
    \langle A(t_1) \rangle _W = 1
\end{equation}

\begin{equation} \label{eq:Bt1}
    \langle B(t_1) \rangle _W = -1
\end{equation}

\begin{equation} \label{eq:At2}
    \langle A(t_2) \rangle _W = 0
\end{equation}

\begin{equation} \label{eq:Bt2}
    \langle B(t_2) \rangle _W = 0
\end{equation}

\begin{equation} \label{eq:Ct2}
    \langle C(t_2) \rangle _W = 1
\end{equation}

\begin{equation} \label{eq:At3}
    \langle A(t_3) \rangle = -1
\end{equation}

\begin{equation} \label{eq:Bt3}
    \langle B(t_3) \rangle _W = 1
\end{equation}

Classical, single-time intuition permits each non-zero single-time weak value to return the truth value for the proposition related to the single-time weak values. Then, the particle's trajectory is deduced from the logical combination of multiple propositions. By appealing to this intuition, the weak values are interpreted as follows. The weak value $\langle A(t_1) \rangle _W$ in Eq. \ref{eq:At1} corresponds to the single-time proposition `The particle is in box $A$ at time $t_1$'. As the weak value is positive, this proposition returns the truth value TRUE. The weak value $\langle B(t_3) \rangle _W$ in Eq. \ref{eq:Bt3} is also positive, and so its corresponding proposition, `The particle is in box $B$ at time $t_3$' likewise returns the truth value TRUE. The same argument applies to Eq. \ref{eq:Ct2} where the weak value $\langle C(t_2) \rangle _W$ corresponds to the proposition `The particle is in box $C$ at time $t_2$' which also returns TRUE.

The null weak value $\langle A(t_2) \rangle _W$ in Eq. \ref{eq:At2} corresponds to the proposition `The particle is in box $A$ at time $t_2$' and returns the truth value FALSE. Likewise, the weak value $\langle B(t_2) \rangle _W$ in Eq. \ref{eq:Bt2} corresponds to the proposition `The particle is in box $B$ at time $t_2$' which also returns the truth value FALSE.

The weak value $\langle B(t_1) \rangle _W$ in Eq. \ref{eq:Bt1} corresponds to the proposition `The particle is in box $B$ at time $t_1$' and has a negative outcome. Rather than negating this proposition, the negative weak value is interpreted as a feature of the non-classical behaviour of the particle, suggesting interference effects. This weak value is argued to interfere with the weak value from Eq. \ref{eq:At1}, thus negating both corresponding propositions \cite{aharonov2017case}. This interpretation of the negative weak value involves a logical combination of the two corresponding single-time propositions using the $\land$ operation, where the first proposition has the truth value TRUE, and the second proposition does not have a truth value due to its negativity. The combination forms the complex proposition, `The particle is in box $A$ at time $t_1$ and the particle is in box $B$ at time $t_1$'. This interpretation of the negative weak value assumes that the complex proposition returns the truth value FALSE, even though this conclusion arose from the logical combination of a truth value TRUE and a non-truth value. There thus seem to be some logical inconsistency between the single-time propositions and the complex proposition which renders this interpretation of the negative weak value problematic. This same argument applies to the negative weak value $\langle A(t_3) \rangle$ in Eq. \ref{eq:At3}.

In an effort to maintain logical consistency, we only perform a logical combination of single-time propositions which have truth values, focusing on the positive and null weak values only. In Sec \ref{Subsec:solution}, we will return to the problem of the negative weak values and provide an alternative interpretation of them.

Combining the positive and null weak values is a logical combination of their corresponding single-time propositions. This looks like the following combination: `The particle is in box $A$ at time $t_1$' $\land$ `The particle is in box $A$ at time $t_2$' $\land$ `The particle is in box $B$ at time $t_2$' $\land$ `The particle is in box $B$ at time $t_3$'. Each proposition has truth values TRUE, FALSE, FALSE, TRUE, respectively. Thus, the complex proposition, `The particle is in box $A$ at time $t_1$ and the particle is in box $A$ at time $t_2$ and the particle is in box $B$ at time $t_2$ and the particle is in box $B$ at time $t_3$' returns FALSE. So from this interpretation, it appears that the particle does not follow the continuous trajectory from box $A$ to box $B$ over the time duration $[t_1,t_3]$. Instead, by considering only the positive weak values, the particle seems to disappear from box $A$ after $t_1$ and re-appear in box $B$ at $t_3$, meaning that the particle does not persist across the duration $[t_1,t_3]$, thus implying temporal discontinuity in the particle's trajectory.

Considering $\langle C(t_2) \rangle _W$ as part of the logical combination does not resolve the discontinuity problem. Considering only the positive weak values and their corresponding TRUE propositions, the logical combination of these results in the following TRUE complex proposition: `The particle is in box $A$ at time $t_1$ and the particle is in box $C$ at time $t_2$ and the particle is in box $B$ at time $t_3$'. The trajectory deduced from this complex proposition is discontinuous in space as the particle does not follow the continuous spatial path $A \rightarrow B \rightarrow C$. This discontinuity leads to tension between Assumptions 1 and 2, suggesting a possible problem with the logical combination of single-time propositions.

We consider alternative methods of measuring a continuous particle trajectory. A recent measurement of sequential weak values (SWVs) in the three-box gendankenexperiment suggests a way of recovering the continuity of the particle's trajectory in spacetime \cite{okamoto2023experimentally}. The authors measure the triple-order SWVs of the spatial projection operator across the boxes, spread over three times $t_1$, $t_2$ and $t_3$. They obtain the following results:

\begin{equation} \label{eq:SWV1}
    \langle A(t_1)A(t_2)B(t_3) \rangle _W = 0.42 \pm 0.06
\end{equation}

\begin{equation} \label{eq:SWV2}
    \langle A(t_1)B(t_2)B(t_3) \rangle _W = 0.54 \pm 0.08
\end{equation}

The first SWV Eq \ref{eq:SWV1} investigates the spatial projection operator of the particle across boxes $A$ at $t_1$, $A$ at $t_2$, and $B$ at $t_3$. Considering all three times together in one SWV results in a positive outcome. This SWV is demonstrative of the path from box $A$ to box $B$ over the time duration $[t_1,t_3]$. The second SWV Eq \ref{eq:SWV2} investigates the spatial projection operator of the particle across boxes $A$ at $t_1$, $B$ at $t_2$, and $B$ at $t_3$. Again, considering all three times together in one SWV provides a positive outcome and the SWV demonstrates the path from box $A$ to box $B$ over the time duration $[t_1,t_3]$. But this SWV investigates a different spatial position at $t_2$ --- $B$ rather than $A$ in the first SWV. So, the two SWVs signify a similar path, but each with a different location at $t_2$, thereby exploring both options for the related trajectory. As both SWVs are positive and approximately $0.5$ each, considering them both together indicates with near certainty the particle trajectory $A \rightarrow B$ over the duration $[t_1,t_3]$.

The SWVs related to the particle on this trajectory do not indicate that the particle jumps through space to a location that is space-like separated from it, nor do they indicate that it jumps through time by disappearing at $t_1$ and re-appearing at $t_3$. Thus, the trajectory appears to be continuous in spacetime, unlike the trajectory deduced from single-time weak values for the equivalent scenario.

We thus argue that we ought to use SWVs to deduce particle trajectories, due to their ability to recover continuity and bring about harmony between Assumptions 1 and 2. The following presence criterion reflects this argument, by demanding a pair of overlapping non-zero SWVs to permit a claim of particle presence:

\hfill

\begin{changemargin}{0.5cm}{0.5cm} 

A pair of overlapping $n$-tuple sequential weak values of the spatial projection operator, each yielding non-zero results, permit a claim of particle presence across the locations and duration described by the SWVs.

\end{changemargin}

\hfill

By `overlapping', we refer to the overlap in time between two SWVs: a pair of SWVs are overlapping when each SWV has one component which investigates a different spatial position on the same path at the same time, for example the triple-order SWVs $\langle X(t_1)X(t_2)Y(t_3) \rangle _W$ and $\langle X(t_1)Y(t_2)Y(t_3) \rangle _W$ at $t_2$ each investigate a different spatial position --- $X$ and $Y$ respectively, which lay on the same path as described by each of these SWVs.

The question is: why do SWVs that meet the presence criterion seem to be able to recover trajectory continuity in spacetime, whilst single-time weak values introduce discontinuity?

\subsection{A solution from logic and consistency} \label{Subsec:solution}

We argue that SWVs are able to recover trajectory continuity as they ask a compound question, forming a consistent set of histories. Meanwhile, using single-time weak values to deduce trajectories involves a logical combination of single-time propositions which are not consistent.

Typically, a set of `histories' is a set of propositions, each at a different successive single time. We borrow the concept from the consistent histories interpretation of quantum mechanics \cite{griffiths1984consistent}. Whilst we don't wholly conform to the interpretation, we briefly discuss it in order to understand the concept of histories that we do borrow.

The consistent histories interpretation explains quantum phenomena by denying that the measurement process plays any fundamental role in quantum theory; the probabilities assigned to events are included in the fundamental axioms of the theory and are independent of any measurement procedure. Measurement outcomes are viewed as an event in the history of the system and the probabilities are associated with consistent sets of histories of events, rather than with any single event \cite{griffiths2017quantum}. Consistent histories rejects unicity --- the belief that at any single point in time there is only one true state of the universe \cite{griffiths2003consistent}. Instead, the interpretation embraces the notion of frameworks, arguing that quantum systems can have multiple frameworks where each framework gives the history of possible states at each point in time. Then a consistent framework consists of a family of histories for which all the histories are jointly exhaustive and mutually exclusive \cite{griffiths2003consistent}:

\newtheorem{defn}{Definition}[section]
\begin{definition}[Jointly exhaustive]
  A jointly exhaustive set of histories provides a complete description of all possible outcomes of a quantum system.
\end{definition}

This is such that all the histories in the set \{$H_i$\} sum to the identity $\mathbb{I}$ and are thus jointly exhaustive as they exhaust all possible outcomes.

\begin{definition}[Mutually exclusive]
  A mutually exclusive set of histories ensures that each history in the set describes a different outcome.
\end{definition}

Any two commuting frameworks can have consistent histories, meanwhile frameworks which do not commute cannot have consistent histories. For example, the questions `What is the spin in the $x$-direction?' and `What is the spin in the $z$-direction?' are inconsistent as spin in the $x$-direction and spin in the $z$-direction are non-commuting properties. The questions are not mutually exclusive because the corresponding spin operators $S_x$ and $S_z$ do not commute. The two frameworks corresponding to these two questions and their individual histories thus do not share consistent histories. Meanwhile, the two possible outcomes for a spin measurement in the $x$-direction of a spin-$\frac{1}{2}$ particle are $+\frac{1}{2}$ and $-\frac{1}{2}$. These possible outcomes are jointly exhaustive in the sense that they exhaust all possible $x$-spin states of the particle. They also commute meaning they are mutually exclusive. They thus fulfil the conditions for consistency and therefore share consistent histories, meaning that both histories are jointly realised according to the consistent histories interpretation \cite{griffiths2014new}.

We borrow this definition of consistency to explain the consistency provided by SWVs. However, the concept of histories that we adopt is dissimilar in the sense that SWVs ask compound questions first, before giving answers in the form of propositions, and these questions form consistent histories. We will demonstrate this in the following, arguing that SWVs include within them the evolution of the position state, and that they can thus be used to deduce consistent particle trajectories.

As our argument adopts the notion of histories from the consistent histories interpretation, one rebuttal is to say that it may run into some of the same problems as those facing the consistent histories interpretation. One particular critique refers to the instability of the past that arises from contradictory histories \cite{dowker1996consistent}. For a quantum system in a particular state, there are many different histories that may describe the evolution of the system into that state. The problem is that the interpretation does not tell us how to single out just one history from this set as the actual descriptor of the evolution of the state. This means that under the consistent histories interpretation, it is possible to have multiple contradictory descriptions of the past, making it difficult to reconstruct a coherent description of past events.

By borrowing from the consistent histories interpretation, our approach may also suffer from this instability of past events. For example, the histories indicated by the SWVs measured provide different accounts of events due to their overlap. But this does not lead to any inability to reconstruct these events. Instead, our approach forms a story about past events by suggesting a superposition of the two paths described by these histories. Accepting this non-classical feature thus makes it possible to reconstruct the past, and so our approach does not suffer from this problem.

We now consider the problem of using single-time weak values to deduce particle trajectories. As already established, classical intuition takes a number of weak values measured at successive single times and logically combines these to deduce the trajectory of the associated particle. The measurement of each single-time weak value corresponds to a question, which is answered by a proposition with truth value TRUE or FALSE. A logical combination of single-time weak values is then a logical combination of the truth values of these propositions. This means that the truth values of the single-time propositions are found before performing the logical combination that deduces a proposition about the trajectory. This method of deduction is similar to the typical understanding of consistent histories, as the propositions form a set of histories which are then combined to deduce the trajectory.

For the particle to have a continuous spacetime trajectory, we assume that the position state of the particle is continuously evolving through time. This is such that the answer to a question regarding the position of the particle at a single time should be dependent on the evolution of the position state over a duration, and so each single-time question ought to be interdependent on each other question. But each single-time weak value asks a question at a single, discrete time only, and then writes down an answer. As each answer is only found at a single time, the answers do not address the evolution of the state across the duration, and they erase the dependency of the single-time positions on each other, making the answers inconsistent with each other and with the evolution of the state. The single-time answers are then logically combined to deduce a proposition about the trajectory, which also forgoes the interdependency of the state's evolution as it is formed from a combination of inconsistent single-time answers. As a result, the single-time weak values form an inconsistent set of histories which do not logically combine to produce a correct deduction of the continuous particle trajectory.

In classical physics, there is a clear consistency between questions and answers that allows us to go back and forth between them. For example, we use answers to single-time questions to deduce the answer to a question across a time duration, and vice versa. But in quantum theory, this relationship breaks down due to the erasure of the interdependency from single-time answers. Instead, we consider compound across-time questions to be primary, such that questions regarding the trajectory are answered without any need for a logical combination of answers, avoiding erasure.

For example, the measurement of the two SWVs $\langle A(t_1)A(t_2)B(t_3) \rangle _W$ and $\langle A(t_1)B(t_2)B(t_3) \rangle _W$ that seem to recover trajectory continuity each correspond to a compound question that can be asked of the quantum system. The first SWV arises from the question: `Is the particle in box $A$ at time $t_1$ and box $A$ at time $t_2$, and box $B$ at time $t_3$?'. This is a compound question as it consists of three parts:

\begin{enumerate}
    \item Is the particle in box $A$ at $t_1$ $\land$,
    \item is the particle in box $A$ at $t_2$ $\land$,
    \item is the particle in box $B$ at $t_3$?
\end{enumerate}

Each part of the question has an answer at the associated single time, but combining each part into one compound question with the logical $\land$ operator avoids asking the single time questions, and instead asks a question across the time duration $[t_1,t_3]$. The combination operation $\land$ is performed prior to obtaining any answers, compared to the single-time method whereby answers to the single-time questions are obtained before performing the combination. And so, the SWV includes the evolution of the position state within it. For example, it is possible to rewrite $A(t_2)$ in terms of $A(t_1)$ just by applying the Hamiltonian. As such, the SWV first asks a question regarding the evolution of the position state across a time duration, and it then writes down an answer. This answer corresponds directly to the path being considered. For example, the first SWV asks the compound question above and then receives an answer in the form of a proposition, assigning a truth value to this. The truth value corresponds to the realisation of the path considered by the measurement of the SWV.

This method of deducing trajectories forms a history which we call History A, but rather than this being a history of propositions, here a history is a compound question across a number of successive times. The compound question consists of possible events, each of which is dependent upon the realisation of other events in the history and on the evolution. This notion of a history avoids the problematic combination of answers, only receiving one answer after asking the compound question related to the SWV.

The second SWV $\langle A(t_1)B(t_2)B(t_3) \rangle _W$, also arises from asking a compound question: `Is the particle in box $A$ at time $t_1$ and box $B$ at time $t_2$, and box $B$ at time $t_3$?'. This question is likewise composed of three parts, where again each part is connected by the logical $\land$ operator:

\begin{enumerate}
    \item Is the particle in box $A$ at $t_1$ $\land$,
    \item is the particle in box $B$ at $t_2$ $\land$,
    \item is the particle in box $B$ at $t_3$?
\end{enumerate}

This question also forms a history, which we call History B. The two histories are jointly exhaustive; they are almost identical and only differ in the part of the question over $t_2$. The variation over $t_2$ exhausts all possibilities --- whilst just one of the histories alone contains the entire duration $[t_1,t_3]$, together both histories contain all possible spatio-temporal positions the particle could move through on a trajectory across the duration. Also, the operators are commutative so the sets are orthogonal, satisfying the mutual exclusivity condition for consistency.

\begin{figure}[H]
\centering
\begin{subfigure}{0.3\textwidth}
    \includegraphics[width=\textwidth]{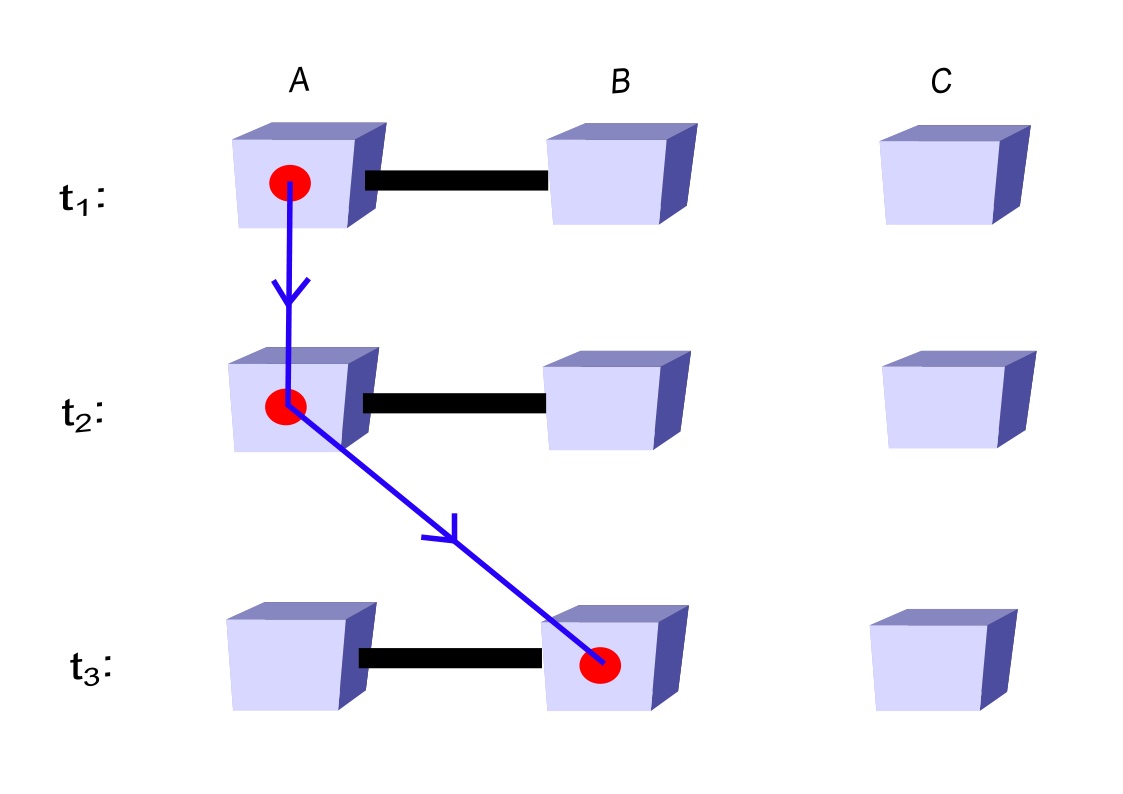}
    \caption{$\langle A(t_1)A(t_2)B(t_3) \rangle _W$}
    \label{fig:first}
\end{subfigure}
\hfill
\begin{subfigure}{0.3\textwidth}
    \includegraphics[width=\textwidth]{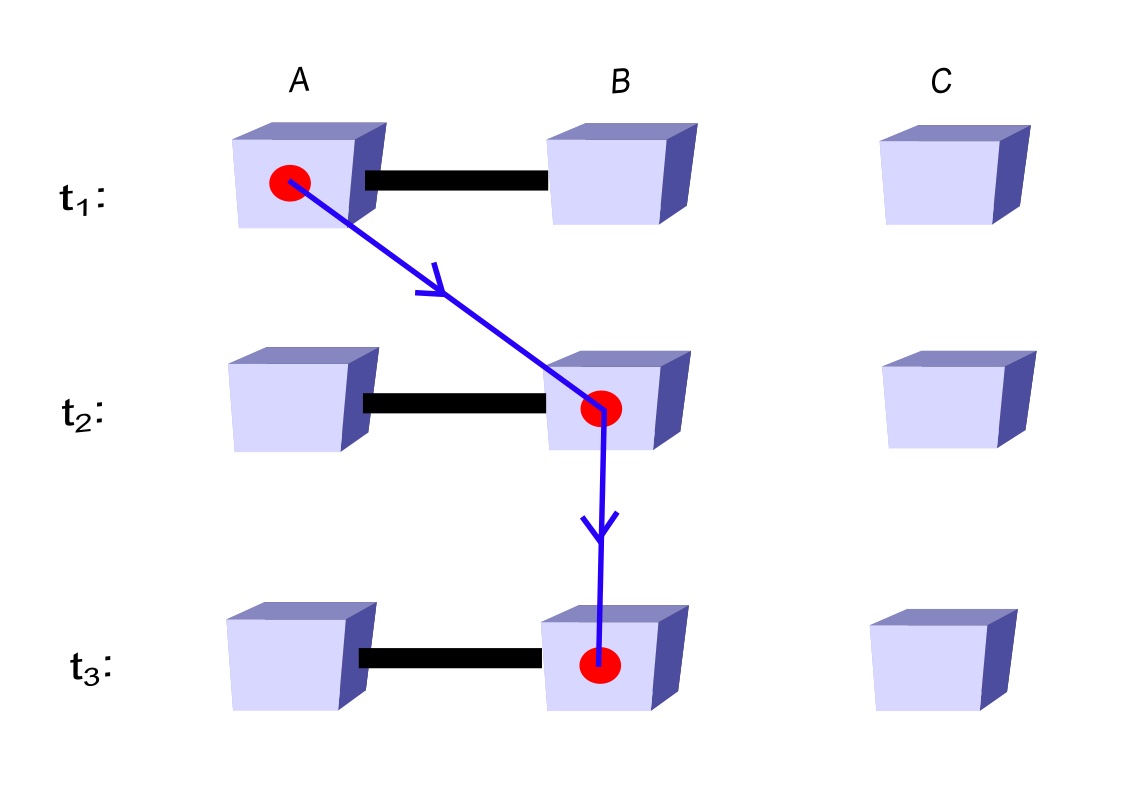}
    \caption{$\langle A(t_1)B(t_2)B(t_3) \rangle _W$}
    \label{fig:second}
\end{subfigure}
\hfill
\begin{subfigure}{0.3\textwidth}
    \includegraphics[width=\textwidth]{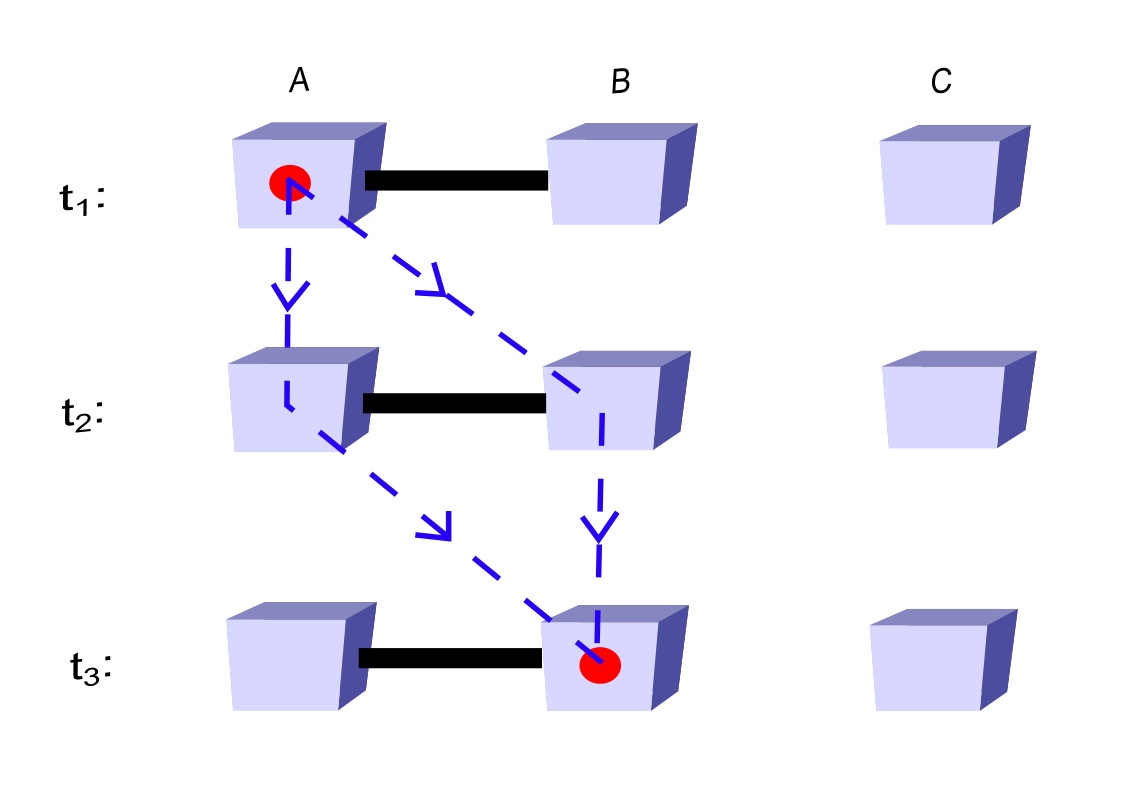}
    \caption{Overlapping SWVs}
    \label{fig:overlap}
\end{subfigure}
        
\caption{The path as defined by SWVs. (a) Demonstrates the path considered by History A. (b) Demonstrates the path considered by History B. (c) Demonstrates the particle's trajectory as superposition of both paths in (a) and (b), such that we know where the particle starts and where it ends up, and that the trajectory between these locations is continuous, but we do not know about the exact position of the particle at any discrete, single time.}
\label{fig:SWVs}
\end{figure}

The pair of histories each ask a compound question, from which a proposition can be deduced and a truth value assigned. Non-zero SWVs return the truth value TRUE. As both SWVs are non-zero, their propositions, `The particle is in box $A$ at time $t_1$ and box $A$ at time $t_2$, and box $B$ at time $t_3$' and `The particle is in box $A$ at time $t_1$ and box $B$ at time $t_2$, and box $B$ at time $t_3$' each return TRUE. The SWVs suggest that the particle is superposed across the two paths described by the corresponding propositions. And these paths are each continuous as they are described by the SWVs which include the continuous evolution of the position state within them. So, the particle appears to be in a superposition of both continuous paths as demonstrated in Fig \ref{fig:SWVs}.

Recalling the negative weak values from Sec \ref{Subsec:paradox}, the problem is that they do not provide either truth value: TRUE or FALSE. So, attempting to perform a logical combination of the negative weak values with the positive and null weak values involves an attempt to combine truth values with non-truth values, which are incompatible. But as there is no need to combine truth values to deduce the trajectory for SWVs, there is no combination of truth values with non-truth values, so this method avoids the combination problem posed by negative weak values.

SWVs recover trajectory continuity by asking across-time questions directly, rather than by deducing the answer to an across-time question from single-time propositions. This result is not surprising, as \cite{fankhauser2021not} critiques weak velocity measurements using single-time weak values, arguing that they fail to provide new empirical evidence for trajectories without assuming the truth of the guidance equation in the de Broglie-Bohm theory. Meanwhile, we have argued that SWVs address the evolution of a particle’s position state across time, and thus they are suitable for deducing trajectories.

We thus argue that quantum theory is about compound questions, not single-time answers, due to the inconsistency in the logical combination of single-time answers to deduce trajectories. By declaring compound across-time questions primary, we prioritise an across-time account of presence, rejecting the single-time intuition. In the following section, we develop this account of presence, and explore some advantages and drawbacks of this account, including the cost of accepting bi-location in order to recover continuity.

\section{Presence from Sequential Weak Values} \label{Sec:acrosspres}

\subsection{Across-time particle presence}\label{analysisofaccount}

As discussed in Sec \ref{Sec:cont}, the classical account of particle presence considers single-time particle positions to be primary, and identified by single-time weak values or eigenvalues. Of course, our classical theories do employ time evolution within them, and as such it is not the theories themselves that embrace the single-time intuition. Rather, it is our classical measurement procedures that support this intuition. Measurements are made at discrete times, and are thus seen as a probing of the continuous time evolution of a system's state, at a discrete time. But this single-time probing leads to measurement outcomes that are based on discrete times. This results in an erroneous reconstruction of the system's evolution via a logical combination of multiple single-time propositions. Consequently, the reconstructed evolution is considered secondary to the single-time measurement outcomes. As such, in classical particle presence, multiple single-time positions of the particle are logically combined to deduce its trajectory, but this trajectory is regarded as secondary to the individual single-time positions from which it is reconstructed.

It is therefore worthwhile to consider an alternate account of particle presence which does not assume the single-time intuition. Whilst the classical, single time account of presence allows presence to be assigned at discrete points in time, we instead restrict claims of presence to cases where an entity persists in spacetime, forming an across-time account of presence.

Persistence is claimed for an entity when that entity exists at multiple times, but there are different definitions for what it means for something to `persist' \cite{lewis1986plurality}. First, an entity \textit{perdures} if it has spatial and temporal parts, meaning that the entity only partially exists at each moment \cite{hales2003endurantism}. Imagine that Alice loses her tennis match and is feeling sad. Alice then goes online and watches some cat videos which cheer her up. Alice therefore has two different properties at these two different moments in time. Given this, it is difficult to ascertain whether Alice persists, as \textit{Sad Alice} becomes \textit{Happy Alice}. The perdurantist deals with this issue by arguing that Alice has different temporal parts at different times; \textit{Sad Alice} and \textit{Happy Alice} are not two separate entities, they are the same entity --- Alice, with different temporal parts, \textit{Sadness} and \textit{Happiness}. Second, an object \textit{endures} if it only has spatial parts, such that the object exists wholly throughout time. The endurantist would deal with Alice's changing emotions by saying that whatever Alice feels \textit{now} is how Alice exists. So if Alice is sad now, then Alice exists as \textit{Sad Alice}, and \textit{Happy Alice} does not exist anywhere in the past, nor in the future. This makes explanations of change difficult for the endurantist, who often concedes to a presentist view of time in order to explain such change.

In our view, particles ought to perdure in spacetime in order to be present. This is because particles undergo change, and perdurantism provides identity whilst accounting for this change. For example, the properties of the particle may change; at one moment in time the particle may have a spin of $+\frac{1}{2}$, and at another moment it may have spin $-\frac{1}{2}$. The perdurantist account allows the particle to persist even given this change in spin. Similarly, experiments such as the quantum Cheshire cat suggest a complex relationship between particles and their properties, but perdurantism provides a means of persistence despite this complexity. In the three-box \textit{gedankenexperiment}, we have considered the changing positions of a particle, which also gives us reason to claim perdurantism, or else endurantism would not be able to explain this change in position without yielding to presentism, which we do not intend to do. Likewise, if we wish to apply this account to entities other than fundamental particles, we run into more problems regarding the change and composition of those entities, and thus perdurance is a safer choice than endurance.

Perdurantists, and thus our across-time account, reject the idea of a privileged present, such that there is no objective present moment moving through time. Instead, our account adopts a four-dimensional view, whereby entities are extended in both the spatial and the temporal dimensions. To say that an entity is present is to say that the entity is extended in time, and thus that it perdures in time, for some duration. This is not to say that the entity must be infinitely extended in time in order for it to ever be present. For example, we could say that a candle flame is present, right now, as it is extended in four dimensions. But if in five minutes a gust of wind extinguishes the flame, then we can no longer say that the flame is present, as it is no longer extended into the future. If we have presence we ought to have absence, and so we see absence as a lack of extension in four dimensions.

This across-time approach to presence manifests in our three-box example as we have seen that SWVs consider the perdurance of the particle in spacetime, whilst the single-time method fails to do so. The SWVs help us to directly deduce the particle's trajectory, without performing a logical combination of single-time positions. We thus demonstrate the relationship between trajectories and single-time positions in our across-time account of presence with the following analogy. 

\begin{figure}
\centering
\begin{subfigure}{0.48\textwidth}
    \includegraphics[width=\textwidth]{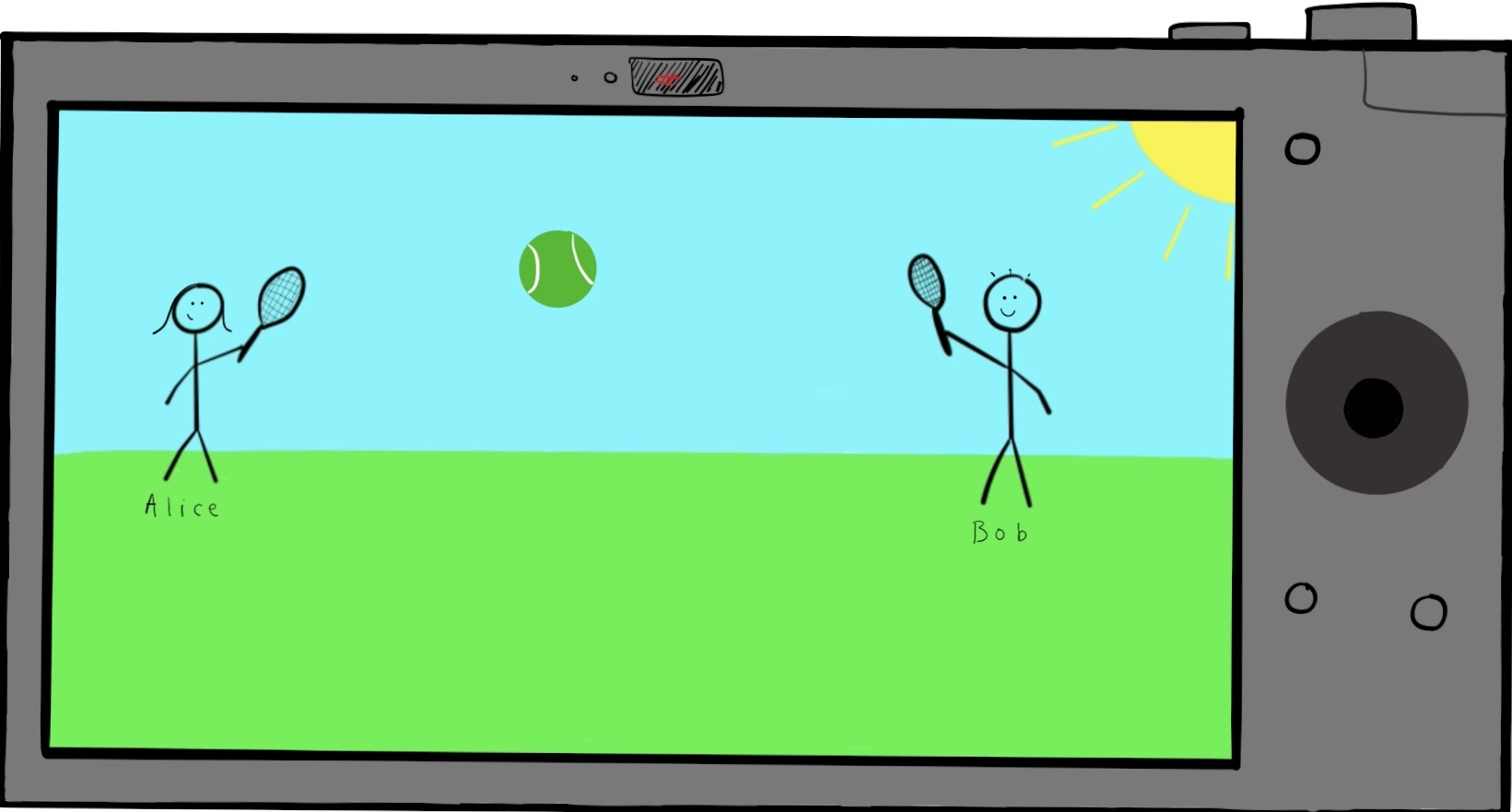}
    \caption{Charlie's perspective}
    \label{fig:spectator}
\end{subfigure}
\hfill
\begin{subfigure}{0.48\textwidth}
    \includegraphics[width=\textwidth]{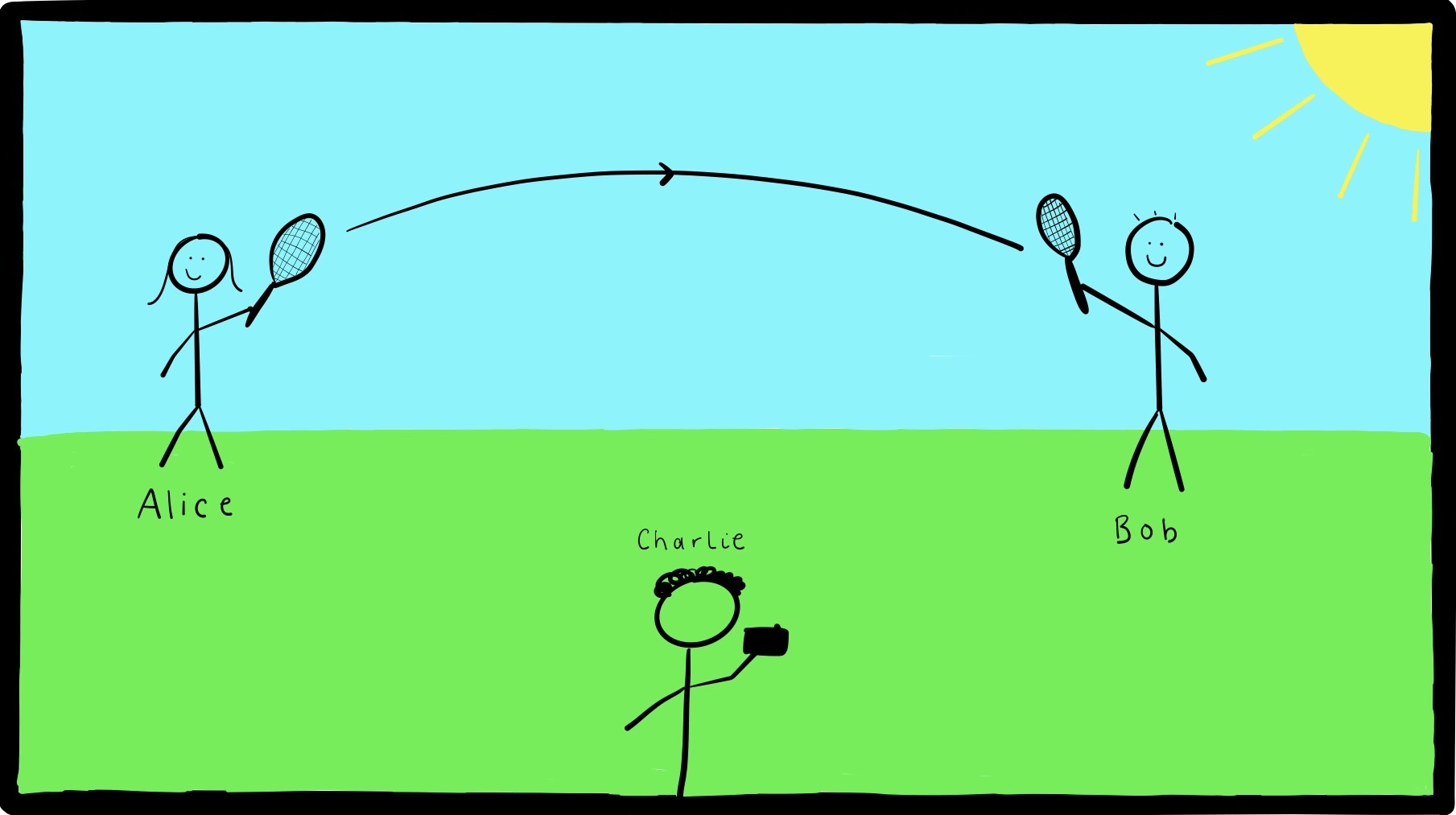}
    \caption{Camera snapshot}
    \label{fig:camera}
\end{subfigure}
\caption{Snapshots of trajectories. (a) From Charlie's perspective, the tennis ball is on a continuously evolving trajectory between Alice and Bob; according to Charlie, its trajectory is not divided into discrete points. (b) Meanwhile, the camera's snapshot shows the tennis ball suspended mid-air in a single position at a single, discrete point in time.}
\label{fig:tennisgame}
\end{figure}

Consider again the analogy of the tennis ball undergoing projectile motion. Figure \ref{fig:tennisgame} demonstrates the approach taken by the across-time account for this analogy. In Fig \ref{fig:spectator} a spectator, Charlie, is watching a game of tennis, and sees the tennis ball on a trajectory between the players, Alice and Bob. From Charlie's perspective, this trajectory is continuous as the position of the tennis ball is continuously evolving through spacetime. Then Charlie takes out a camera and snaps a photograph of the tennis ball on its trajectory. Figure \ref{fig:camera} demonstrates the photograph taken by the camera. The photograph is of the tennis ball in a single spacetime position, which alone does not reveal the continuous evolution of the ball's position, as seen by Charlie. The photograph is just a snapshot of the ball on its trajectory, and although we may be able to logically infer that the ball follows a continuous trajectory due to our experience in the world, the snapshot alone does nothing to confirm this.

As the snapshot does not capture the evolution of the ball's spatial position across time, it does not address its perdurance in spacetime and thus does not say anything about the presence of the ball. So, the snapshot is secondary to Charlie's perspective, as presence is assigned to Charlie's perspective rather than to that of the camera.

Particle presence can be thought of in a similar way. Trajectories are analogous to Charlie's perspective as they each directly address the evolution of the position state, and thereby address the perdurance in spacetime. Meanwhile, single-time propositions are analogous to the snapshot taken by the camera as they arise from the single-time measurement procedure, probing the evolution of the position state at a discrete time, but not including the evolution within them and thus not addressing the perdurance of the particle in spacetime. So, like the snapshot, single-time propositions also tell us anything about presence. Particle trajectories are thus considered primary in the across-time account, whilst single-time positions can be thought of as a snapshot of a trajectory.

In the case that it is necessary to find the position of a particle at a single-time, one ought to do so by measuring SWVs. For example, if we wanted to know whether or not the particle was in box $A$ at time $t_2$, measuring $\langle \hat{A}(t_2) \rangle _W$ is not appropriate as the single-time weak value does not say anything about the perdurance of the particle in spacetime, and thus does not correspond to any proposition about the presence of the particle. Instead, we should measure overlapping SWVs that correspond to a trajectory where the particle may pass through $A$ at $t_2$. If these SWVs are non-zero then we can say that the particle follows a continuous trajectory through $A$ across the duration being considered. However, it is not possible to be certain that the particle is in $A$ at the discrete time $t_2$ as the SWV allows only compound, across-time propositions to be made. If one wishes to make a proposition about the position of the particle at $t_2$, then it is a single-time proposition which is purely a snapshot of the trajectory, and does not assign particle presence.

Embracing this across-time approach comes at the cost of accepting bi-location. The SWV measurement procedure resolves the discontinuity problem, and we obtain a superposition of continuous particle trajectories as a result. So, we ought to be very clear about what the across-time account of particle presence means, including what we gain from this account, and equivalently what we may have to sacrifice. To do so, we characterise our across-time account of particle presence through a revision of the HRL conditions for classical particle presence. We eliminate single-time intuition and thereby arrive at the following conditions for across-time particle presence:

\begin{itemize}
    \item [(i*)] A particle's trajectory is continuous in time.
    \item [(ii*)] A particle's trajectory is continuous in space.
    \item [(iii*)] A particle only interacts with objects or fields local to its spacetime trajectory.
    \item [(iv*)] A particle's property must be on the same spacetime trajectory that the particle is on.
\end{itemize}

Several of the HRL conditions are revised in a way that eliminates the single-time intuition. HRL condition (i) becomes across-time condition (i*) after being reformulated to make it clear that this condition demands continuity in time. The single-time intuition has been eliminated by removing the idea that the particle is located in a single position at a discrete time, and replacing this with the assertion that the particle's \textit{trajectory} is continuous. This same method applies to HRL condition (iii) which becomes across-time condition (ii*), specifying that particle trajectories are expected to be continuous in space. Condition (ii*) does not assume single-time intuition as it assumes that trajectories are primary, rather than being built from single-time statements regarding a particle's position. Across-time condition (iii*) replaces HRL condition (iv) by specifying a `spacetime trajectory' rather than a `spacetime position', also eliminating single-time intuition. Likewise, HRL condition (vi) becomes across-time condition (iv*). As a result, we assert the same message as HRL that properties ought not be delocalised from particles, but without asserting the problematic single-time intuition.

HRL condition (ii) is removed as in the across-time account it is reasonable for a particle to probabilistically be on more than one path. For example, the SWVs $\langle A(t_1)A(t_2)B(t_3) \rangle _W$ and $\langle A(t_1)B(t_2)B(t_3) \rangle _W$ define two different possible paths that the particle could take, and each has an approximately equivalent probability, such that the particle is considered to be in a superposition of both boxes $A$ and $B$ \cite{aharonov2017case}. Under our account, each possible path is equally realised and so we avoid making assertions such as HRL condition (ii). This means that superposition is a key feature of the across-time account of particle presence. We do not claim that particles are always bi-located, just that they \textit{can be}. This is a departure from the classical account of particle presence, as now a particle can be present in the sense that it is following two different trajectories over the same time duration.

We also remove HRL condition (v) as this condition demands that a particle is located in the same region that its trajectory is located, but with the revised continuity conditions (i*) and (ii*), this is obvious. For if a particle is on a continuous trajectory, then it cannot be considered to be separate from its trajectory and thus it must be located within the same region as its trajectory is. So, condition (v) is certainly a feature of the across-time account, but we do not see it necessary to specify so in the form of a condition, as it is naturally included in our intuition when considering the across-time account.

We have demonstrated that the revised conditions are sufficient for claiming particle presence, now we will show that they are also necessary. Consider a particle moving between two locations, $X$ and $Y$ across some duration $[t_i,t_f]$. Condition (i*) and (ii*) are consistent with the assumption of a continuous spacetime manifold, and also ensure that the particle perdures in spacetime between $X$ and $Y$ across $[t_i,t_f]$. If the particle does not meet either of these first two conditions then it does not perdure in spacetime and thus a proposition regarding its presence will return FALSE.

Condition (iii*) is necessary if locality is to be maintained. If one admits to non-local correlations then this condition is not necessary as the particle could in that case interact with non-local objects and fields. But as our approach makes no specific comment on locality, we argue that condition (iii*) is necessary in order to retain some classical features. The particle must then only interact with objects and fields that are local to the trajectory that it is on between locations $X$ and $Y$, across the duration $[t_i,t_f]$. However, we accept that the necessity of this condition is contingent on one's stance on locality.

We argue that condition (iv*) is also necessary to claim particle presence. Certain effects such as the quantum Cheshire cat suggest the disembodiment of a property from its corresponding particle, and can even demonstrate an apparent exchange of properties between two particles \cite{das2020can}. We argue that this effect results in the indistinguishability of particles, making it difficult to determine whether a particle perdures and thereby whether it is present. So, in order to be present under the across-time account, condition (iv*) is necessary.

This is not to say that these revised assumptions are not challenged by any other quantum phenomena. What we insist is that if we adopt the across-time account of presence, these are the conditions that we can recover, but we must pay the price of losing some of our other conditions (HRL (ii) and (v)), meaning that while we may be able to recover continuity, we cannot maintain all of our classical intuition. In our view, this provides us with a way to deal with the discontinuity that we began with, realigning Assumptions 1 and 2, and thus this is a cost we are willing to pay, as recovering continuity provides us with a way to identify and track particles, being a pragmatic tool for providing a clear narrative of particle spatiotemporal evolution.

This account posits quasi-classical particle trajectories; whilst these trajectories maintain certain features of classicality, they also exhibit non-classical features such as the possibility of superposition and the rejection of the single-time intuition about measurements. But in doing so, this account brings about cohesion between Assumptions 1 and 2, providing explanatory power and pragmaticism, and thus we argue that this move to quasi-classicality is justified.

\subsection{Solving the nested interferometer paradox}
\label{solving}

The across-time account may be extended to other cases of trajectory discontinuity in order to make sense of the paradoxes. We demonstrate the resolution of the apparent discontinuity in the nested interferometer paradox by calculating SWVs that meet the presence criterion.

\begin{figure}
    \centering
    \includegraphics[scale=0.3]{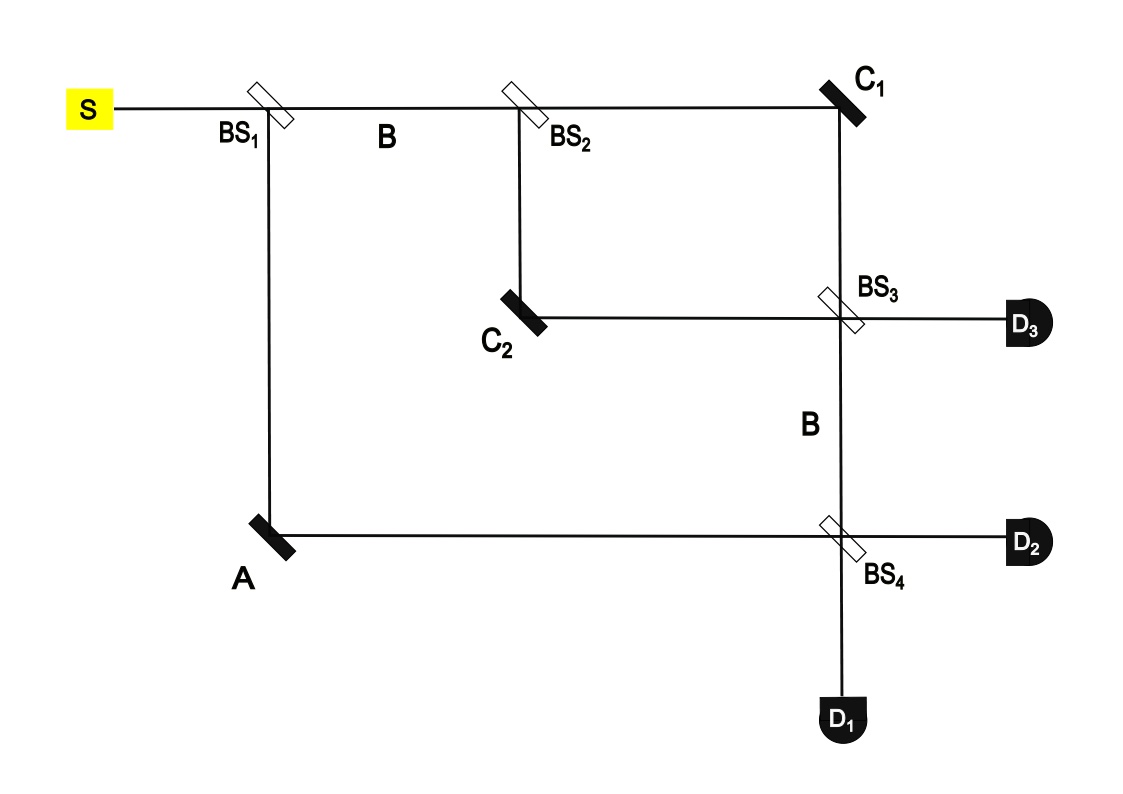}
    \caption{Nested MZI: A Mach-Zehnder interferometer is nested in the right arm $B$ of a second Mach-Zehnder interferometer. Pictured are the photon source $S$, arms $A$, $B$, $C_1$, and $C_2$, beam-splitters $BS_1$, $BS_2$, $BS_3$, and $BS_4$, and detectors $D_1$, $D_2$, and $D_2$. The black lines demonstrate the possible paths the photon could take through the interferometer.}
    \label{fig:nestedMZI}
\end{figure}

The experimental set-up involves a Mach-Zehnder interferometer (MZI) nested within the right arm of a second MZI as seen in Fig \ref{fig:nestedMZI} \cite{danan2013asking}. The left arm is denoted $A$, the right arm $B$, and the nested interferometer is $C$, where the upper and lower arms of the nested interferometer are denoted $C_1$ and $C_2$. It is expected that a photon that enters the outer MZI from the source $S$ becomes superposed over arms $A$ and $B$ after passing through the first beam-splitter $BS_1$. Then, by travelling arm $B$, the photon enters the nested interferometer $C$ via the second beam-splitter $BS_2$ and is superposed over arms $C_1$ and $C_2$.

Calculating the single-time weak values in each arm gives the following outcomes \cite{vaidman2013past,vaidman2014tracing}:

\begin{equation}\label{eq:intA}
    \langle \hat{P}_A(t) \rangle_W =1
\end{equation}

\begin{equation}\label{eq:intC1}
    \langle \hat{P}_{C_1}(t) \rangle_W =\frac{1}{2}
\end{equation}

\begin{equation}\label{eq:intC2}
    \langle \hat{P}_{C_2}(t) \rangle_W =-\frac{1}{2}
\end{equation}

\begin{equation}\label{eq:intB}
    \langle \hat{P}_B(t) \rangle_W =0,
\end{equation}

where $\hat{P}$ is the projection operator for each arm $A$, $C_1$, $C_2$, and $B$ respectively. Applying single-time intuition to these weak values results in the following interpretation. Equation (\ref{eq:intA}) corresponds to the proposition, `The photon is in arm $A$ at time $t$' with truth value TRUE. Equation (\ref{eq:intC1}) corresponds to the proposition, `The photon is in arm $C_1$ at time $t$' with truth value TRUE, and Eq (\ref{eq:intC2}) corresponds to the proposition, `The photon is in arm $A$ at time $t$' also with truth value TRUE. Note that the second and third weak values are $\frac{1}{2}$ each. According to the single-time intuition, these non-zero weak values assign presence of the photon in both arms $C_1$ and $C_2$ of the nested interferometer simultaneously. A logical combination of both of the corresponding propositions of $\langle \hat{P}_{C_1}(t) \rangle_W$ and $\langle \hat{P}_{C_2}(t) \rangle_W$ deduces an additional proposition, `The photon is in arm $C_1$ and in arm $C_2$ at time $t$' which returns TRUE, suggesting the superposition of the photon across both arms. Meanwhile, the null weak value (\ref{eq:intB}) corresponds to the proposition `The photon is in arm $B$ at time $t$' with truth value FALSE.

The proposition `The particle is in arm $A$ and in arm $C_1$ and $C_2$ at time $t$' is deduced from the application of the logical combination operator $\land$. This proposition returns TRUE, suggesting that the photon is superposed across arms $A$, $C_1$, and $C_2$. However, the lack of presence in arm $B$ deduced from the null weak value results in a discontinuous trajectory as the particle is unable to follow a continuous trajectory to $C$ without passing through $B$. Note that it is not the superposition that is primarily problematic to us, rather it is the discontinuous trajectory that we have a problem with, due to the tension this generates between Assumptions 1 and 2, and due to our desire for continuity for pragmatic purposes.

Two-state vector formalism provides one explanation of this discontinuity, by arguing that the particle’s trajectory is defined by the overlap in the states evolving forward in time from the pre-selection, and backward in time from the post-selection \cite{vaidman2014tracing}. By post-selecting for the case where the photon ends up detected by $D_2$, the overlap in the forward and backward evolving states correspond to the non-zero single time weak values, Eq (\ref{eq:intA}, (\ref{eq:intC1}, and (\ref{eq:intC2}. As a result, one explanation is that the discontinuity is not problematic, as it is just a feature of these overlapping states. However, our intention is to see whether it is possible to recover continuity here, as it provides a coherent narrative of the photon’s evolution.

We calculate the SWVs for this paradox in order to investigate the evolution of the state across the time duration. In doing so, we avoid the erroneous combination of single-time propositions and directly measure the evolution of the particle's position state, thereby measuring the perdurance of the particle in spacetime. Using the pre-selection $\ket{\psi}=\frac{1}{\sqrt{3}}(\ket{A}+\ket{B}+\ket{C})$, and post-selection $\ket{\phi}= \frac{1}{\sqrt{3}} (\ket{A}+\ket{B}-\ket{C})$, we find:

\begin{equation}\label{eq:AAA}
    \langle \hat{P}_{A}(t_1) \hat{P}_{A}(t_2) \hat{P}_{A}(t_3) \rangle _W = 1
\end{equation}

\begin{equation}\label{eq:C1C1C2}
    \langle \hat{P}_{C_1}(t_1) \hat{P}_{C_1}(t_2) \hat{P}_{C_2}(t_3) \rangle _W = 0
\end{equation}

\begin{equation}\label{eq:C1C2C2}
    \langle \hat{P}_{C_1}(t_1) \hat{P}_{C_2}(t_2) \hat{P}_{C_2}(t_3) \rangle _W = 0
\end{equation}

The intention of Eq \ref{eq:AAA} is to investigate only arm $A$ over the duration [$t_1,t_3$] and thus its overlapping pair is not required. This SWV asks the question `Is the photon in arm $A$ across the duration [$t_1,t_3$]?' which returns TRUE. From our presence criterion, we confirm the perdurance of the photon in arm $A$ over this duration, so we can claim that the particle is present in arm $A$ over this duration in the sense of the across-time account of presence that we have developed. Equations \ref{eq:C1C1C2} and \ref{eq:C1C2C2} are overlapping pairs. Each corresponds to the respective questions: `Is the photon is in arm $C_1$ at time $t_1$ and arm $C_1$ at time $t_2$ and arm $C_2$ at time $t_3?$' and `Is the photon is in arm $C_1$ at time $t_1$ and arm $C_2$ at time $t_2$ and arm $C_2$ at time $t_3?$'. These questions each return FALSE, and thus from the presence criterion, the photon is not considered to be present in arm $C$ across the duration [$t_1,t_3$].

The lack of presence in arm $B$ is unproblematic as we also deduce a lack of presence in $C$, so we avoid the discontinuity. Rather, the photon follows a continuous trajectory only on arm $A$ across the duration [$t_1,t_3$]. Not only does this approach eliminate the discontinuous trajectories, but it also avoids the admission of a superposition of trajectories, providing an almost classical description of the trajectory events inside the interferometer. The only remaining divergence from classicality is the departure from single-time intuition. But, as the across-time account provides a measurement procedure that directly addresses the evolution of the state, we argue that this account provides a much more accurate measurement scheme, and thus the rejection of single-time intuition is not a loss.

\section{Further Research Directions}
\label{further}

\begin{figure}
    \centering
    \includegraphics[scale=0.4]{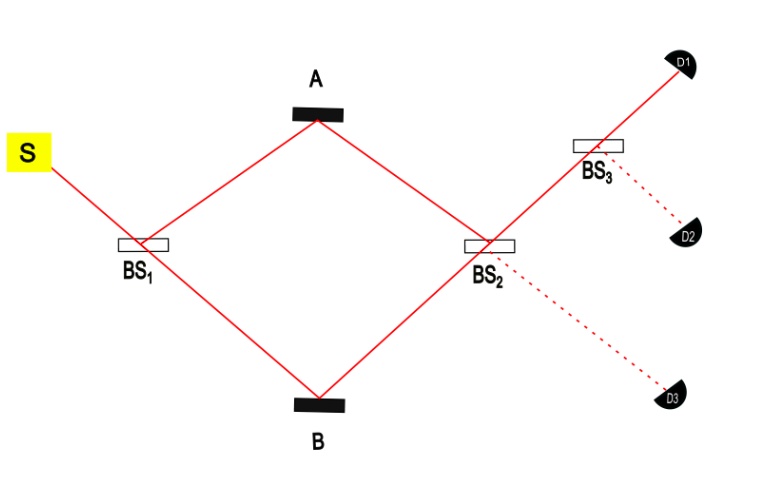}
    \caption{Experimental set-up for the demonstration of the kinematic quantum Cheshire cat effect. A beam of pre- and post-selected neutrons traverse through paths $A$ and $B$ of an interferometer. Introducing a magnetic field to path $A$ does not affect the spin of the neutrons located there. However, introducing the field to path $B$ does have this effect.}
    \label{fig:Cheshirecat}
\end{figure}

We have shown that the across-time account resolves some discontinuity paradoxes which arise from the erroneous assumption of single-time intuition. It is thus reasonable to expect that this account may be extended to additional discontinuity paradoxes, in hope that it resolves these also.

One paradox is the quantum Cheshire cat effect, which exploits weak measurements to demonstrate the delocalisation of property from particle \cite{guryanova2012complete,ibnouhsein2012twin,aharonov2013quantum}. A beam of neutrons are pre-selected in state $\ket{\uparrow_x}$, and they enter an interferometer on a superposition on both possible paths $A$ and $B$, as shown in Fig \ref{fig:Cheshirecat}. A spin-rotator is introduced to path $B$ which has the effect of flipping the spin of the neutrons in path $B$ from the pre-selected state $\ket{\uparrow_x}$ to the state $\ket{\uparrow_z}$. Path $A$ is then defined by the spin state $\ket{\uparrow_x}$, while path $B$ is defined by the spin state $\ket{\uparrow_z}$. By post-selecting in state $\ket{\uparrow_x}$, the output signal is restricted to neutrons which travelled path $A$ only. Thus, when a weak magnetic field is introduced to path $A$, it is expected that the field will interact with the neutrons in path $A$ and alter their spin as a result of this interaction. But via an observation of an interference effect at the output, it is shown that the spin of the neutrons in the post-selected state is only impacted when the field is on path $B$ \cite{denkmayr2014observation}. This result suggests a disembodiment of the polarisation from the neutron, such that the neutron follows path $A$, while its polarisation follows path $B$. This interpretation leads to confusion regarding the relationship between particles and their properties, specifically how we ought to define particles if not in reference to their properties.

We suggest that our across-time account may be able to make sense of this disembodiment effect. The paradox arises from a measurement of single-time weak values, and thus we suggest that a measurement of SWVs could be performed to trace the path followed by the neutron and its polarisation. It is then possible to ask a compound question regarding the neutron's polarisation over a time duration, possibly rectifying discontinuity between the neutron and its polarisation. If so, then a more well-defined understanding of particles and particle presence is attained, as questions concerning the disembodiment of properties from particles are avoided. However, this method involves the extension of our across-time judgment of weak values to operators other than the spatial projection operator. So, one open question asks whether the measurement of weak values of the polarisation operator should be restricted to single-time weak values, or else are SWVs necessary for determining the spin of a particle? We suggest that in the case that the spin-value of a particle is used to deduce the position of the particle, the across-time judgment may apply to the related weak values, such as is the case for the Cheshire cat effect. We suggest this as an interesting route for further investigation.

In this article, we have worked under the assumption of a continuous, fixed background spacetime, where events can be associated with specific spacetime points. But an additional open question concerns how we can speak about presence in diffeomorphism-invariant frameworks. In frameworks such as general relativity, spacetime is no longer a fixed background stage upon which events unfold and are associated with specific spacetime locations. As spacetime points do not have intrinsic physical significance, there is no privileged set of spacetime points where measurements ought to be performed. So associating presence to measurements linked to specific spacetime points is problematic, and we are left without a clear account of presence within diffeomorphism-invariant frameworks. It is thus necessary to explore alternative ways of describing the perdurance of entities over time. Thus, a possible future research direction is to consider whether we can maintain an across-time account of presence that does not rely on fixed spacetime coordinates, which may involve a more relational or holistic understanding of presence.

One way to investigate presence without a fixed background spacetime is through an application of the across-time account to indefinite causality. Typically, causality is understood via operations that are performed in a fixed, well-defined order. For example, operation A occurs before operation B, $A>B$. However, by exploiting quantum superposition, it is possible to consider causal orders which are not well-defined, such that the causal order is placed in a superposition of both $A>B$ and $B>A$. The process matrix framework provides a description of processes that are not constrained by a well-defined order, relating the inputs and outputs of different parties without imposing a strict order of operations \cite{oreshkov2012quantum}.

To further clarify this framework, we could trace the path of the system as it navigates through this indefinite order of operations, possibly gaining insight into how the system's behaviour conveys the indefinite causal structure. We suggest doing so via a measurement of SWVs, which ask a compound question regarding the perdurance of the system between operations. This approach may provide insight into the relationships between operations that defy a well-defined order, helping to clarify the indefinite causal order framework. While in this article we have shown that entities perdure through a local time, i.e. within localised causal frameworks where definite causal order is applicable, this SWV approach to indefinite causality may suggest that perdurance can be demonstrated without fixed causal relationships, within a global time. We doubt that perdurance in local time is defined similarly to perdurance in global time, due to the lack of definition of global time, so we suggest that a distinction ought to be made between the local and global definitions of perdurance. This once again challenges classical intuitions about temporality, raising a question: is perdurance defined `in time', or is time defined by perduring entities? It is interesting to consider the case where time emerges from the perdurance of an entity, which may provide a means of understanding presence without a fixed temporal background.

\section{Conclusion} \label{Sec:conc}

The analysis of presence in this article suggests that the classical, single-time intuition about presence may be incorrect. Our measurement procedures na\"ively assume this intuition, and as a result, they do not consider the evolution of a system in its entirety, thus leading to problems such as discontinuous particle trajectories. Instead, we have argued that to denote presence, measurements must take on an across-time character. By this, we mean that presence should be understood as a property which is defined across time, and that our measurement procedures ought to reflect this. We suggest that presence should not be measured as a property materialising at discrete times, and thus urge the avoidance of making propositions about presence at single times. To denote presence, we argue that one should measure SWVs as this method involves asking a compound question regarding the trajectory, and addresses the evolution of the position state across a time duration. This method forms a consistent set of histories, recovering the continuity of the particle's trajectory. This across-time analysis suggests an alternative metaphysical account of presence and may also be extended to develop a more intuitive account of the presence of macroscopic objects.

The account of presence we have developed herein is influenced by Bergson's concept of \textit{la dur\'ee} (duration). Bergson introduced the theory of time as duration as an inherently continuous and interconnected process which cannot be broken down into distinct points \cite{bergson1911essai}. Bergson thus critiques the notion of `spatialised time', where time is treated as a series of static, discrete, homogeneous points, with one point causally influencing the next \cite{durie1999introduction}. He contends that time ought to instead be viewed as continuously evolving, such that the present is constantly approaching, and as soon as it approaches it has already passed \cite{Bergson1910-BERTAF-3}. Both Bergson’s view and our own reject a purely discrete or presentist view of time: in Bergson’s account, time is irreducible to isolated moments; it is a seamless flow. In our account, the continuity of temporal parts resists the notion of strict discrete presence. Bergson’s durée captures the intrinsic, indivisible nature of continuity, which resonates with our attempt to define quantum particles as perduring entities in spacetime. Our across-time account aligns the measurement procedure with a more Bergsonian view of time and presence, addressing the evolution of the particle's position in spacetime, resulting in an account of presence which is more closely aligned with this phenomenological view.

Our analysis in this article has shown that classical, single-time intuition about particle presence is flawed, leading to tension between Assumptions 1 and 2. Embracing an across-time approach via SWVs brings about cohesion between these assumptions and leads to a perspective on presence in which perdurance is prioritised. This approach not only clarifies what we can say about particle presence and trajectories, but also opens avenues for further research in cases where the perdurance status of an entity is unclear.

\section*{Acknowledgments}

I extend my gratitude to Alexei Grinbaum, who provided valuable comments on several drafts of this article, and with whom I had many helpful discussions about the ideas presented in the article. This research was funded, partially, by l’Agence Nationale de la Recherche (ANR), project ANR-22-CE47-0012.

\section*{Declarations}

The authors declare no competing interests.


\end{document}